\documentclass[11pt,a4paper]{article}
\usepackage{jcapmod}
\usepackage{booktabs}
\usepackage[english]{babel}
\usepackage{amsmath,amssymb,amsbsy,amstext, amsthm}
\usepackage{hyperref}
\usepackage{graphicx}
\usepackage{amsfonts}
\usepackage{amssymb}
\usepackage{upgreek}
\usepackage{exscale,relsize}
\usepackage[makeroom]{cancel}
\usepackage{soul}

\allowdisplaybreaks[1]

\newcommand{\nn}{\nonumber}

\newcommand{\be}{\begin{equation}}
\newcommand{\ee}{\end{equation}}
\newcommand{\bea}{\begin{eqnarray}}
\newcommand{\eea}{\end{eqnarray}}
\newcommand{\bk}{\boldsymbol{k}}
\newcommand{\bp}{\boldsymbol{p}}
\newcommand{\bq}{\boldsymbol{q}}
\newcommand{\bx}{\boldsymbol{x}}
\newcommand{\br}{\boldsymbol{r}}
\newcommand{\bd}{\boldsymbol{d}}

\makeatletter
\newlength{\apb@width}
\newcommand{\autoparbox}[2][c]{\settowidth{\apb@width}{#2}\parbox[#1]{\apb@width}{#2}}

\makeatother

\numberwithin{equation}{section}

\def\beq{\begin{equation}}
\def\eeq{\end{equation}}

\def\bea{\begin{eqnarray}}
\def\eea{\end{eqnarray}}

\begin{document}

\begin{titlepage}

\setcounter{page}{1} \baselineskip=15.5pt \thispagestyle{empty}
\begin{flushright}
DESY 14-215
\end{flushright}
\bigskip\

\begin{center}

{\fontsize{20}{28}\selectfont  \sffamily \bfseries 
On Soft Limits of Large-Scale Structure\\ 
\vskip 0.5cm 
Correlation Functions} 

\end{center}

\vspace{0.2cm}

\begin{center}
{\fontsize{13}{30}\selectfont  Ido~Ben-Dayan, Thomas~Konstandin, Rafael~A. Porto and  Laura~Sagunski}
\end{center}

\begin{center}
\textsl{Deutsches Elektronen-Synchrotron DESY, Theory Group, D-22603 Hamburg, Germany}
\end{center}

\vspace{1.2cm}
\hrule \vspace{0.3cm}
\noindent {\sffamily \bfseries Abstract} \\[0.1cm]

We study soft limits of correlation functions for the density and velocity fields in the theory of structure formation. First, we re-derive the (resummed) consistency conditions at unequal times using the eikonal approximation. These are solely based on symmetry arguments and are therefore universal. Then, we explore the existence of equal-time relations in the soft limit which, on the other hand, depend on the interplay between soft and hard modes. We scrutinize two approaches in the literature: The time-flow formalism, and a background method where the soft mode is absorbed into a locally curved cosmology. The latter has been recently used to set up (angular averaged) `equal-time consistency relations'. We explicitly demonstrate that the time-flow relations and `equal-time consistency conditions' are only fulfilled at the linear level, and fail at next-to-leading order for an Einstein de-Sitter universe. While applied to the velocities both proposals break down beyond leading order, we find that the `equal-time consistency conditions' quantitatively approximates the perturbative results for the density contrast. Thus, we generalize the background method to properly incorporate the effect of curvature in the density and velocity fluctuations on short scales, and discuss the reasons behind this discrepancy. We conclude with a few comments on practical implementations and future directions.
\vskip 10pt
\hrule

\vspace{0.6cm}

\bigskip

\end{titlepage}

\tableofcontents

\newpage

\section{Introduction}

Soft limits, that link $(n+1)$-point and $n$-point correlators of the density perturbations in the theory of structure formation, have recently received significant attention. The main appeal of these relations (for unequal times) is that they are solely based on the assumption of a single-field inflationary background, providing the seed for the initial conditions, together with the diffeomorphism invariance of General Relativity. Therefore, they lead to (quite generally) non-perturbative statements about the system on short scales that serve as a probe of basic aspects of the theory at hand~\cite{Maldacena:2011nz, Sherwin, baldauf2, Kehagias:2013yd, Peloso:2013zw, Creminelli:2013mca, Peloso:2013spa, Creminelli:2013poa, Creminelli:2013nua}. This becomes a very powerful tool in the context of using the forthcoming large-scale structure (LSS) surveys to test single-field inflation as a theory of initial conditions for the seed of structure, as well as the equivalence principle in gravitational theories, especially since fluctuations enter the non-linear regime at small redshift.

Various approaches have been used to derive soft limits for correlation functions in LSS.\footnote{Soft limits of inflationary correlation functions have also been extensively studied in the literature, e.g. \cite{fgp,paolo1,dansoft,Gold,Lam}.} As~stated, these relations are most meaningful for correlation functions at different times. For equal-time correlators they become degenerate, in the sense that they vanish at leading order in $q$, with $q$ being the soft (or long) mode.\footnote{Technically speaking, when dealing with the density field (as opposite to the potential) the leading-order term scales like $q^{-1}$.} To extract information about equal-time correlators, one therefore has to study next-to-leading order (NLO) effects, where dynamical information, as opposite to gauge artifacts in General Relativity, start to become important~\cite{baldauf2, Sherwin, Creminelli:2013cga}. It is then relevant to determine whether at equal times one may be able to write down expressions that are still valid even when the short modes are deep in the non-linear regime. In fact, allegedly non-perturbative relations in the soft limit at equal times have been recently advocated in the literature~\cite{Kehagias:2013paa,Valageas:2013zda}.\vskip 4pt

The purpose of this paper is thus twofold.  In section \ref{sec:2}, we first (re-)derive the consistency conditions for the soft (squeezed) limit of density and velocity correlators at \textit{unequal times}. Here we exploit the compact notation of the fluid equations in the Eulerian representation of perturbation theory, which simultaneously includes both fields, e.g \cite{Scoccimarro:2000zr,Bernardeau:2001qr}. One of the crucial aspects in the derivation of the unequal-time relation is the factorization of soft and hard modes, which can be resummed into an eikonal phase.

Next, we explore under which circumstances relations between \textit{equal-time} correlators in the soft limit may exist beyond a perturbative treatment of the hard modes, when the coupling between long and short fluctuations becomes important. For this purpose, we compute in section \ref{sec:SPT} the three-point function (or bispectrum) at NLO in the soft limit as a benchmark for comparison of different methods.
We study two different approaches that have appeared in the literature. The reader may choose to concentrate on one or the other without disturbing the flow of the paper. 

In section \ref{sec:4}, we study the time-flow formalism \cite{Pietroni:2008jx} which relies on applying a `closure,' or truncation, approximation to a hierarchical set of evolution equations. For the case of the bispectrum, we show that the connected piece of the four-point function (or trispectrum), often neglected in the literature, plays an important role in assessing the validity of an equal-time relation. In general, only perturbative statements may be derived in the time-flow approach for a given truncation. In section \ref{sec:5}, we study the  implementation of a map, discussed in \cite{baldauf2,Sherwin}, between dynamics on short scales within a flat Friedmann-Robertson-Walker (FRW) universe in the presence of a long-wavelength perturbation and a locally curved FRW background. 
In the context of $N$-body simulations, this map was exploited in the so-called `separate universe' approach and used to
compute the power spectrum response function~\cite{Li:2014sga, Wagner:2014aka}~\footnote{For other non-perturbative approaches involving the halo model see~\cite{Li:2014sga, Chiang:2014oga}.}. 
Furthermore, this equivalence was applied in \cite{Kehagias:2013paa,Valageas:2013zda} to propose angular averaged soft-limit relations for correlation functions at equal times. As we argue, the expression derived in \cite{Kehagias:2013paa,Valageas:2013zda} does carry information that is non-perturbative in the short modes, for example coming from the `displacement' term (or equivalently the eikonal phase). However, the part that accounts for the growth of structure cannot be formally extended beyond leading order in perturbation theory, not even in an Einstein de-Sitter (EdS) universe. In spite of this, while the proposal of  \cite{Kehagias:2013paa,Valageas:2013zda} fails when applied to the velocity beyond linear order, we find that it produces quantitatively accurate results for the density contrast compared to standard perturbation theory (SPT). For example, for the bispectrum of density fluctuations at one-loop order the error is only a few percent. We generalize the background method to properly incorporate the effect of curvature in the density and velocity fluctuations on short scales, which we show react differently (by a factor of order one) to the presence of a local curvature, and discuss the reasons behind this discrepancy. We conclude in section \ref{sec:disc}, with a discussion on the accuracy of the `equal-time consistency conditions' of \cite{Kehagias:2013paa,Valageas:2013zda} and future directions. 

\section{Correlation Functions at Unequal Times in the Soft Limit}\label{sec:2}

\subsection{Fluid equations}

If we ignore deviations from the perfect fluid approximation, which are required to account for the imprint of hard modes on the long-distance scales \cite{eft1,eft2}, the non-linear fluid equations of cosmological perturbation theory (the continuity, Euler and Poisson equations) can be expressed in a compact form by writing the matter density contrast $\delta$
and the divergence of the velocity field in Fourier space as doublet $\psi_{a}\!\left(\boldsymbol{k},\eta\right)$
with $a\in\left\{ 1,2\right\} $,
\begin{equation}
\psi_a\! \left(\boldsymbol{k},\eta\right)
\equiv \left(\begin{array}{c}
\delta\!\left(\boldsymbol{k},\eta\right)\\
\Theta(\bk,\eta)
\end{array}\right)\,,\label{eq: fields}
\end{equation}
with $\Theta (\bk,\eta) \equiv -\boldsymbol{\nabla}\cdot\boldsymbol{v}\left(\boldsymbol{k},\eta\right)/\mathcal{H}$.
Thereby, the conformal time $\tau$ has been replaced by the time
variable $\eta\equiv\ln a(\tau)$ in terms of the
scale factor $a(\tau)$ and $\mathcal{H}\equiv\partial\ln
a\!\left(\tau\right)/\partial\,\tau = a \, H$
denotes the conformal expansion rate. 
In general background cosmologies, the fluid equations then read \cite{Scoccimarro:2000zr,Bernardeau:2001qr}
\begin{equation}
\partial_{\eta}\psi_{a}\!\left(\boldsymbol{k},\eta\right)=-\Omega_{ab}
\!\left(\boldsymbol{k},\eta\right)\,\psi_{b}\!\left(\boldsymbol{k},
\eta\right)+ \gamma_{abc}\!\left(\boldsymbol{k},\boldsymbol{-p},
-\boldsymbol{q}\right)\,\psi_{b}\!\left(\boldsymbol{p},\eta\right)\psi_{c}
\!\left(\boldsymbol{q},\eta\right)\,,\label{eq:  fluid equations}
\end{equation}
using the convention that repeated indices are summed and
integration over internal momenta has to be performed whenever the
vertex function
$\gamma_{abc}\!\left(\boldsymbol{k},\boldsymbol{p},\boldsymbol{q}\right)$
appears. The only independent, non-vanishing, elements of
$\gamma_{abc}\!\left(\boldsymbol{k},\boldsymbol{p},\boldsymbol{q}\right)$
with $a,b,c\in\left\{ 1,2\right\} $, 
\begin{equation}
\begin{alignedat}{1}\gamma_{121}\!\left(\boldsymbol{k},\boldsymbol{p},
\boldsymbol{q}\right)=\gamma_{112}\!\left(\boldsymbol{k},\boldsymbol{q},
\boldsymbol{p}\right) &
=\frac{1}{2}\,\delta^{\textnormal{D}}\!\left(\boldsymbol{k}+\boldsymbol{p}
+\boldsymbol{q}\right)\cdot\alpha\!\left(\boldsymbol{p},\boldsymbol{q}\right)\,,\\
\gamma_{222}\!\left(\boldsymbol{k},\boldsymbol{p},\boldsymbol{q}\right) &
=\,\delta^{\textnormal{D}}\!\left(\boldsymbol{k}+\boldsymbol{p}
+\boldsymbol{q}\right)\cdot\beta\!\left(\boldsymbol{p},\boldsymbol{q}\right)\,,
\end{alignedat}
\label{eq: gamma elements}
\end{equation}
 arise as products of the Dirac delta-distribution, denoted by
$\delta^{\textnormal{D}}\!\left(\boldsymbol{k}+\boldsymbol{p}+\boldsymbol{q}
\right)$,
and the functions 
\begin{equation}
\alpha\!\left(\boldsymbol{p},\boldsymbol{q}\right)=\frac{\left(\boldsymbol{
p}+\boldsymbol{q}\right)\cdot\boldsymbol{p}}{p^{2}}\,,
\qquad\beta\!\left(\boldsymbol{p},\boldsymbol{q}\right)=\frac{
\left(\boldsymbol{p}+\boldsymbol{q}\right)^{2}\boldsymbol{p}\cdot\boldsymbol{q}}
{2\, p^{2}q^{2}}\,,
\end{equation}
which couple different modes of density and velocity perturbations.\\

The dependence of the fluid equations \eqref{eq:  fluid equations}
on the cosmological model is encoded in the matrix $\Omega_{ab}$. For example, for the
simplest case of a flat, pure dark matter cosmology with EdS
background, it is given by  
\begin{equation}
\Omega_{ab}=\left(\begin{array}{cc}
\;\;\,0 & -1\\
-\frac{3}{2}\,\Omega_{m} & \;\;\,1+\frac{1}{\mathcal{H}^{2}}\frac{\partial\mathcal{H}}{\partial\tau}
\end{array}\right)=\left(\begin{array}{cc}
\;\;\,0 & -1\\
-\frac{3}{2} & \;\;\,\frac{1}{2}
\end{array}\right)\,.
\label{eq: Omega_ab-1}
\end{equation}
Generalization to other cosmologies is straightforward, e.g. \cite{Pietroni:2008jx}. However, in order to perform explicit computations, we will use \eqref{eq: Omega_ab-1} as a working example, in particular when including higher-order perturbative corrections. In this case, the linear propagator, which describes the time evolution of $\psi_a\! \left(\boldsymbol{k},\eta\right)$ at the linear level, reads
\begin{equation}
g_{ab}\!\left(\eta,\eta'\right)=\left[B\,e^{\left(\eta-\eta'\right)}
+A\,e^{-\frac{3}{2}\left(\eta-\eta'\right)}\right]_{ab}
\theta\!\left(\eta-\eta'\right)\, , \label{eq: g_ab EdS}
\end{equation}
where $\theta\!\left(\eta\right)$ denotes the Heaviside step function
and
\begin{equation}
B=\frac{1}{5}\left(\begin{array}{cc}
3 & 2\\
3 & 2
\end{array}\right)\,,\qquad A=\frac{1}{5}\left(\begin{array}{cc}
\;\;\,2 & -2\\
-3 & \;\;\,3
\end{array}\right)\,.
\end{equation}

\subsection{The eikonal approximation}

The derivation of consistency relations in the squeezed limit is relatively straightforward for unequal times. Since soft and hard modes evolve independently at leading order in $q$, the soft effects can be resummed, yielding an eikonal phase~\cite{Bernardeau:2011vy, Jain:1995kx}.\footnote{See also \cite{Blas:2013bpa, Sugiyama:2013gza} for related discussions.} For the fluctuations of the 
hard modes $\bk$, with $k \gg q$, one finds (we suppress indices in what follows for simplicity)
\be
\label{eq:factor}
\psi({\bk}, \eta) \simeq {\rm exp}\left[{\int^{\eta} d\eta' \int_{\bp}^{\Lambda_L} \, \frac{\bk \cdot \bp }{p^2} \,\Theta_L(\bp,
\eta')}\right]
\, \times \, \psi_S({\bk}, \eta) \, ,
\ee
where $\int_{\bf p} \equiv \int \frac{d^3 \bp}{(2\pi)^3}$. Here, $\psi_S({\bk},\eta)$ denotes the fluctuations of short scales, including interactions
of the type short-short, and $\Theta_L(\bp,\eta)$ are the long-scale modes of the velocity
field, in units of the conformal Hubble parameter according to its definition in \eqref{eq: fields}. We have also introduced a cutoff $\Lambda_L$ to emphasize the integral is performed over soft momenta. Moreover, at leading/linear order we have (for the growing mode) $\Theta_L(\bp,\eta) = \delta_L(\bp,\eta)$. Therefore, the impact of the long modes comes in the form of an exponential that involves the linear density field. In particular, the effect of soft physics completely factorizes at this order. Using the expression in \eqref{eq:factor}, one can readily derive the consistency relations of large-scale structure at unequal times, noticing that
\bea
\left< \psi_L(\bq, \eta_q)\, \psi(\bk_1, \eta_1) \cdots \psi(\bk_n, \eta_n) \right>'
&\xrightarrow{q \to 0}&
\left< \psi_L(\bq, \eta_q) \, {\rm exp}\left[{\sum_i \int^{\eta_i} d\eta'_i \int_{\bp}^{\Lambda_L}  \, \frac{\bk_i \cdot
\bp}{p^2} \,\delta_L(\bp,\eta'_i)}\right]
\right>'\nn \\
&&  \times \, \left< \psi_S(\bk_1, \eta_1) \cdots \psi_S(\bk_n, \eta_n)  \right>' \,, 
\eea
and
\bea
\left<  \psi(\bk_1, \eta_1) \cdots \psi(\bk_n, \eta_n) \right>'
&\xrightarrow{q \to 0}&
\left<  {\rm exp}\left[{\sum_i \int^{\eta_i} d\eta'_i \int_{\bp}^{\Lambda_L}  \, \frac{\bk_i \cdot
\bp}{p^2} \,\delta_L(\bp,\eta'_i)}\right]
\right>' \nn \\
&&  \times \, \left< \psi_S(\bk_1, \eta_1) \cdots \psi_S(\bk_n, \eta_n)  \right>' \,, 
\eea
following the evaluation of the cumulants
\beq
\frac{\left< \psi_L(\bq, \eta_q) \, {\rm exp}\left[{\sum_i \int^{\eta_i} d\eta'_i \int_{\bp}^{\Lambda_L} \, \frac{\bk_i \cdot
\bp}{p^2}\, \delta_L(\bp,\eta'_i)}\right]
\right>'}{\left<  {\rm exp}\left[{\sum_i \int^{\eta_i} d\eta'_i \int_{\bp}^{\Lambda_L} \, \frac{\bk_i \cdot
\bp}{p^2} \,\delta_L(\bp,\eta'_i)}\right]\right>'} = 
- P_L(q, \eta_q) \sum_i \frac{D(\eta_i)}{D(\eta_q)} \frac{\bk_i \cdot \bq}{q^2}\,, \label{cummulant}
\eeq
where the linear power spectrum $P_L(q, \eta)\equiv D^{2} (\eta)\, P_{0}(q)$ is defined in terms of the linear growth factor $D(\eta)$ and the initial power spectrum $P_{0}(q)$. The angular brackets denote ensemble averages, while the prime~$\langle\ldots \rangle'$ indicates that the momentum conserving delta-function has been removed.
This result reproduces the previously derived relations, e.g. \cite{Creminelli:2013poa}, including in addition the velocity field (see \eqref{eq: fields}). Furthermore, it can be easily extended to account for different background cosmologies. 
\vskip 1pt
Notice that this derivation did not require more than the leading-order factorization of long modes in the squeezed limit. However, when $\eta_q = \eta_i$ for all $i$, the right-hand side in \eqref{cummulant} vanishes at leading order in $q$ due to momentum conservation, such that a calculation to second order becomes unavoidable to obtain the behavior in the soft-$q$ limit at equal times. This is a crucial point in determining the validity of relations between correlation functions beyond linear order in such case.

\section{The Bispectrum at Next-to-Leading Order} \label{sec:SPT}

In order to assess the validity of consistency relations between correlation functions at \textit{equal times,} we compute the NLO contribution to the power spectrum and bispectrum of density perturbations in SPT. We will then explore, as a first non-trivial check, the existence of a soft-limit connection between the bispectrum and power spectrum beyond the linear approximation. Note also that SPT predictions should agree with the exact solution within its realm of validity.\footnote{The loop expansion in SPT requires techniques such as the effective field theory framework  for LSS in order to properly account for the imprint of the short-distance physics. Therefore integrals that appear in SPT computations need to be regularized by introducing counter-terms \cite{eft1,eft2,left,Baldauf:2014qfa,Carrasco:2013sva,Leo,Pajer, Dan}. This is even more relevant when dealing with the velocity field, which is a composite operator \cite{left,eft2,Pajer}. Nevertheless, one can always choose fictitious initial conditions for the power spectrum such that SPT converges quickly. The comparisons here may be then understood as mathematical statement at the level of the integrands, while judiciously choosing initial conditions such that the integrals are dominated by the modes well within the perturbative regime.}

To set up our notation, we introduce the two- and three-point correlation functions in terms of the
power spectrum $P_{ab}\!\left(k,\eta\right)$ and bispectrum $B_{abc}\!\left(\boldsymbol{k},\boldsymbol{q,}\boldsymbol{p},\eta\right)$ by defining
\begin{equation}
\begin{alignedat}{1}\langle\psi_{a}\!\left(\boldsymbol{k},\eta\right)\psi_{b}
\!\left(\boldsymbol{q},\eta\right)\rangle &
\equiv\delta^{\textnormal{D}}\!\left(\boldsymbol{k}+\boldsymbol{q}\right)P_{ab}
\!\left(k,\eta\right)\,,\\
\langle\psi_{a}\!\left(\boldsymbol{k},\eta\right)\psi_{b}\!\left(\boldsymbol{q},
\eta\right)\psi_{c}\!\left(\boldsymbol{p},\eta\right)\rangle &
\equiv\delta^{\textnormal{D}}\!\left(\boldsymbol{k}+\boldsymbol{q}+\boldsymbol{p
}\right)B_{abc}\!\left(\boldsymbol{k},\boldsymbol{q,}\boldsymbol{p},\eta\right)\,.
\end{alignedat}
\end{equation}
Moreover, we write the linear power spectrum as $P^{L}_{ab}(k,\eta) = u_a u_b \, P_L(k,\eta)$ with growing mode initial conditions $u_a=(1,1)$. Thus, at leading order the relation between the angular averaged bispectrum in the soft limit and the power spectrum is given by~\cite{Sherwin}
\beq
B^{L}_{111}\!\left(\boldsymbol{k},-\boldsymbol{q},
\boldsymbol{q}-\boldsymbol{k},\eta\right)^{\textnormal{av}} 
\xrightarrow{q \to
0} 
 \, P_{L}\!\left(q,\eta\right)
\biggl( \frac{47}{21} - \frac13 \, k~\partial_k
\biggr)P_L\!\left(k,\eta\right)   \label{eq:squeezedBlinear}\, ,
\eeq

At NLO, or one-loop order, the bispectrum may be obtained in a straightforward manner using the standard techniques. However, the resulting expressions are too lengthy to allow for a meaningful analytic comparison. For ease of use, we then restrict ourselves to the limit where the loop momentum, $l$, is much larger than the external momenta, $k$ or $q$. In this limit the angular averaged squeezed bispectrum may be thus written in the form  
\begin{eqnarray}
B_{111}^{\textnormal{\rm 1-loop}}\!\left(\boldsymbol{k},-\boldsymbol{q},\boldsymbol{q}-\boldsymbol{k},\eta\right)^{\textnormal{av}} & \simeq & P_{L}\!\left(q,\eta\right)\frac{k^{2}}{\pi^{2}} \biggl[ \left(\alpha\, k\,\partial_{k}+\beta\right)P_{L}\!\left(k,\eta\right) \! \int\! dl\, l^{2}\left(\frac{P_{L}\!\left(l,\eta\right)}{l^{2}}\right) \nonumber \\
 &  & \qquad \qquad \quad \;+\gamma \,k^{2} \!
\int\! dl\, l^{2}\left(\frac{P_{L}^{2}\!\left(l,\eta\right)}{l^{4}}\right) \biggr]\,,
\end{eqnarray}
with some numerical factors $\alpha, \beta, \gamma$. These coefficients can be obtained from the one-loop computation of the bispectrum in SPT after taking the squeezed limit and angular average, and we get

\be
\alpha^{\rm SPT} = \frac{61}{1890} \simeq 0.032  \, ,\quad
\beta^{\rm SPT} = - \frac{3719}{13230} \simeq -0.281 \, ,\quad
\gamma^{\rm SPT} =\frac{515}{5292} \simeq 0.097 \, .\quad  \label{4.2}
\ee
We collect a few more details on the SPT computation in appendix \ref{appBispectrum}. For completeness we also include the one-loop contribution to the power spectrum of density/velocity fluctuations in the limit of large loop momentum, which is given by
\bea
\label{eq:SPT_P_1loop}
P_{ab}^{\rm 1-loop}(k,\eta)  &=& 
\left(\begin{array}{cc}
\frac{9}{196} & \frac{19}{588}\\
\frac{19}{588} & \frac{61}{980}\\ 
\end{array}\right)
\frac{k^4}{\pi^2} \int dl \,l^2 \left(\frac{P^2_L(l,\eta)}{l^4}\right) \nn \\
&& - 
\left(\begin{array}{cc}
\frac{61}{630} & \frac{25}{126}\\
\frac{25}{126} & \frac{3}{10}\\ 
\end{array}\right)
\frac{k^2 \, P_L(k,\eta)}{\pi^2} \int dl \,l^2 \left(\frac{P_L(l,\eta)}{l^2} \right)\,.
\eea
In the following sections, we will use these values as a benchmark for evaluating attempts to extend the result in \eqref{eq:squeezedBlinear}.

\section{Correlation Functions at Equal Times I: Time-Flow Approach \label{sec:4}}
 
In this section, we study the time-flow approach, introduced in \cite{Pietroni:2008jx}, to set up correlation functions in the squeezed limit. We will show that soft-limit relations may be derived, however, we will explicitly demonstrate for the case of the bispectrum that these are only valid at leading order. The reader may skip this section on a first read of the manuscript.\\

The time-flow approach consists of multiplying the fluid equations \eqref{eq:  fluid equations} by the fluctuations and take the statistical average. The $\eta$-evolution
of the correlation functions can then be written~as
\begin{equation}
\begin{alignedat}{1}\partial_{\eta}\langle\psi_{a}\psi_{b}\rangle &
=-\Omega_{ac}\langle\psi_{c}\psi_{b}\rangle-\Omega_{bc}\langle\psi_{a}\psi_{c}
\rangle\\
 &
\quad+\gamma_{acd}\langle\psi_{c}\psi_{d}\psi_{b}\rangle+\gamma_
{bcd}\langle\psi_{a}\psi_{c}\psi_{d}\rangle\,,\\
\vphantom{} & \vphantom{}\\
\partial_{\eta}\langle\psi_{a}\psi_{b}\psi_{c}\rangle &
=-\Omega_{ad}\langle\psi_{d}\psi_{b}\psi_{c}\rangle-\Omega_{bd}\langle\psi_{a}
\psi_{d}\psi_{c}\rangle-\Omega_{cd}\langle\psi_{a}\psi_{b}\psi_{d}\rangle\\
 &
\quad+\gamma_{ade}\langle\psi_{d}\psi_{e}\psi_{b}\psi_{c}\rangle+
\gamma_{bde}\langle\psi_{a}\psi_{d}\psi_{e}\psi_{c}\rangle+\gamma_{cde}
\langle\psi_{a}\psi_{b}\psi_{d}\psi_{e}\rangle\,,\\
\vphantom{} & \vphantom{}\\
\partial_{\eta}\langle\psi_{a}\psi_{b}\psi_{c}\psi_{d}\rangle & =\ldots\;.\end{alignedat}
\label{eq: flow equation 1}
\end{equation}
To improve readability, we have omitted the momentum and time
dependence of the correlation functions. All fields have to be evaluated
at the same value of $\eta$. The procedure above produces an infinite hierarchy of evolution equations. Hence, the usefulness of the time-flow approach relies on finding a suitable `closure approximation', as we will review next.

\subsection{Closure approximation}

If we express the four-point correlation functions in terms of the
power spectrum
and the trispectrum, 
%
\bea
 \langle\psi_{a}\!\left(\boldsymbol{k},\eta\right)\psi_{b}\!\left(\boldsymbol{q},
\eta\right)\psi_{c}\!\left(\boldsymbol{p},\eta\right)\psi_{d}\!\left(\boldsymbol
{w},\eta\right)\rangle  &\equiv& \delta^{\textnormal{D}}\!\left(\boldsymbol{k}+\boldsymbol{q}
\right)\delta^
{\textnormal{D}}\!\left(\boldsymbol{p}+\boldsymbol{w}\right)P_{ab}\!\left(k,
\eta\right)P_{cd}\!\left(p,\eta\right) \nn \\
 &+& \delta^{\textnormal{D}}\!\left(\boldsymbol{k}+\boldsymbol{p}\right)\delta^
{\textnormal{D}}\!\left(\boldsymbol{q}+\boldsymbol{w}\right)P_{ac}\!\left(k,
\eta\right)P_{bd}\!\left(q,\eta\right)\nn \\
&+&\delta^{\textnormal{D}}\!\left(\boldsymbol{k}+\boldsymbol{w}\right)\delta^
{\textnormal{D}}\!\left(\boldsymbol{q}+\boldsymbol{p}\right)P_{ad}\!\left(k,
\eta\right)P_{bc}\!\left(q,\eta\right)\nn \\
&+&\delta^{\textnormal{D}}\!\left(\boldsymbol{k}+\boldsymbol{p}+\boldsymbol{q
}+\boldsymbol{w}\right)Q_{abcd}\!\left(\boldsymbol{k},\boldsymbol{q},\boldsymbol
{p},\boldsymbol{w},\eta\right)\,,
\label{eq: correlators}
\eea
the closure approximation at this level consists of neglecting the trispectrum, namely setting $Q_{abcd}\equiv 0$. This then allows us to express the four-point correlation function in terms of power spectra. The flow equations \eqref{eq: flow equation 1}
thus form a \emph{closed} system, which can be formally solved as
\begin{equation}
\begin{alignedat}{1}P_{ab}\!\left(k,\eta\right) &
=g_{ac}\!\left(\eta,\eta_0\right)g_{bd}\!\left(\eta,\eta_0\right)
P_{cd}\!\left(k ,
\eta_0\right)\\
 & \quad+\int_{\eta_0}^{\eta}\! d\eta'\,\int\! d^{3}q\,
g_{ae}\!\left(\eta,\eta'\right)g_{bf}\!\left(\eta,\eta'\right)\\
 &
\quad \quad \times \bigl[\gamma_{ecd}\!\left(\boldsymbol{k},-\boldsymbol{q},
\boldsymbol{q}-\boldsymbol{k}\right)B_{fcd}\!\left(\boldsymbol{k},-\boldsymbol{q
},\boldsymbol{q}-\boldsymbol{k},\eta'\right)\\
 &
\quad\quad\quad \,+\gamma_{fcd}\!\left(\boldsymbol{k},-\boldsymbol{q},\boldsymbol{q}
-\boldsymbol{k}\right)B_{ecd}\!\left(\boldsymbol{k},-\boldsymbol{q},\boldsymbol{
q}-\boldsymbol{k},\eta'\right)\bigr]\,,\\
\end{alignedat}
\label{eq: P_ab}
\end{equation}
and
\begin{equation}
\begin{alignedat}{1}B_{abc}\!\left(\boldsymbol{k},-\boldsymbol{q},\boldsymbol{q
}
-\boldsymbol{k},\eta\right) 
&= g_{ad}\!\left(\eta,\eta_0\right)g_{be}\!\left(\eta,\eta_0\right)g_{cf}
\!\left(\eta,\eta_0\right)B_{def}
\!\left(\boldsymbol{k},-\boldsymbol{q},\boldsymbol{q}-\boldsymbol{k},
\eta_0\right)\\
 & \quad+2\,\int_{\eta_0}^{\eta}\!
d\eta' g_{ad}\!\left(\eta,\eta'\right)g_{be}\!\left(\eta,
\eta'\right)g_{cf}\!\left(\eta,\eta'\right)\\
 &
\quad\quad\times\bigl[\gamma_{dgh}\!\left(\boldsymbol{k},-\boldsymbol{q},
\boldsymbol{q}-\boldsymbol{k}\right)P_{eg}\!\left(q,\eta'\right)P_{fh}
\!\left(\bigl|\boldsymbol{q}-\boldsymbol{k}\bigr|,\eta'\right)\\
 &
\quad\quad\quad\,+\gamma_{egh}\!\left(-\boldsymbol{q},\boldsymbol{q}-\boldsymbol{k},
\boldsymbol{k}\right)P_{fg}\!\left(\bigl|\boldsymbol{q}-\boldsymbol{k}
\bigr|\eta'\right)P_{dh}\!\left(q,\eta'\right)\\
 &
\quad\quad\quad\,+\gamma_{fgh}\!\left(\boldsymbol{q}-\boldsymbol{k},\boldsymbol{k},
-\boldsymbol{q}\right)P_{dg}\!\left(k,\eta'\right)P_{eh}\!\left(q,
\eta'\right)\bigr]\,,
\end{alignedat}
\label{eq: B_abc}
\end{equation}
where $\eta_0$ corresponds to the initial time.

\subsection{The soft limit of the bispectrum}

Provided the closure approximation holds, one may derive an angular averaged  consistency relation for the soft limit $q \to 0$ of the bispectrum $B_{abc}\!\left(\boldsymbol{k},-\boldsymbol{q},\boldsymbol{q}-\boldsymbol{k}
,\eta\right)^{\textnormal{av}}$, and in principle for general classes of cosmologies. We take Gaussian
initial conditions such that the initial bispectrum vanishes. Under these assumptions the expression in \eqref{eq: B_abc}
simplifies to
\begin{equation}
\begin{alignedat}{1}B_{abc}\!\left(\boldsymbol{k},-\boldsymbol{q},\boldsymbol{q}
-\boldsymbol{k},\eta\right) & =2\,\int_{\eta_0}^{\eta}\!
d\eta' g_{ad}\!\left(\eta,\eta'\right)g_{be}\!\left(\eta,
\eta'\right)g_{cf}\!\left(\eta,
\eta'\right)\\
 &
\quad\times\bigl[\gamma_{dgh}\!\left(\boldsymbol{k},-\boldsymbol{q},\boldsymbol
{
q}-\boldsymbol{k}\right)P_{eg}\!\left(q,\eta'\right)P_{fh}
\!\left(\bigl|\boldsymbol{q}-\boldsymbol{k}\bigr|,\eta'\right)\\
 &
\quad\quad\,+\gamma_{egh}\!\left(-\boldsymbol{q},\boldsymbol{q}-\boldsymbol{k},
\boldsymbol{k}\right)P_{fg}\!\left(\bigl|\boldsymbol{q}-\boldsymbol{k}\bigr|,
\eta'\right)P_{dh}\!\left(k,\eta'\right)\\
 &
\quad\quad\,+\gamma_{fgh}\!\left(\boldsymbol{q}-\boldsymbol{k},\boldsymbol{k},
-\boldsymbol{q}\right)P_{dg}\!\left(k,\eta'\right)P_{eh}\!\left(q,
\eta'\right)\bigr] \, .
\end{alignedat}
\label{eq: B_abc-3}
\end{equation}
Next, we expand all quantities in \eqref{eq: B_abc-3} which depend
on the difference between the soft and the hard modes,
$\bigl|\boldsymbol{q}-\boldsymbol{k}\bigr|$, in a perturbative series up to first order in $q$. Inserting the series expansion of
 the power spectrum,
\begin{equation}
P_{ab}\!\left(\bigl|\boldsymbol{q}-\boldsymbol{k}\bigr|,\eta\right) 
\simeq P_{ab}\!\left(k,\eta\right) - q\,\mu \,\partial_{k}P_{ab}\!\left(k,
\eta\right)+\mathcal{O}\!\left(q^{2}\right)\,,
\label{eq: P_ab series}
\end{equation}
with $\mu \equiv \left(\boldsymbol{q}\cdot\boldsymbol{k}
\right)/\left(qk\right)$, we then obtain\footnote{Notice that the second term in (\ref{eq: B_abc-3}) is
linear in $q$ and may be ignored.}
\bea
\label{eq:squeezedB}
&& 
B_{abc}\!\left(\boldsymbol{k},-\boldsymbol{q},\boldsymbol{q}
-\boldsymbol{k},\eta\right) = 
2\,\int_{\eta_0}^{\eta}\!
d\eta' \, g_{ad}\!\left(\eta,\eta'\right)g_{be}\!\left(\eta,
\eta'\right)g_{cf}\!\left(\eta,
\eta'\right) \,  P_{eh}\!\left(q,\eta'\right)\nn \\
&& \times  \, \, \biggl[ 
\gamma_{dhg}\!\left(\boldsymbol{k},-\boldsymbol{q},\boldsymbol{q}-\boldsymbol{k}\right) 
\Bigl( P_{fg}\!\left(k, \eta'\right) - \mu \,
q~\partial_k P_{fg}\!\left(k,\eta'\right)
\Bigr)
+
\gamma_{fgh}\!\left(\boldsymbol{q}-\boldsymbol{k},\boldsymbol{k},
-\boldsymbol{q}\right) P_{dg}\!\left(k,
\eta'\right) 
\biggr] \, .\nn \\
\eea
Furthermore, in the limit $q \to 0$ the bracket becomes
\bea
&& \biggl[ \delta_{h1} \Bigl( 
M^A_{dg}\,  P_{gf}(k,\eta') + P_{dg}(k,\eta') \,  M^A_{gf} 
\Bigr)  \nn \\ \,  &&   + \,\delta_{h2} \Bigl(
M^B_{dg}\, P_{gf}(k,\eta') + P_{dg}(k,\eta') \, M^B_{gf} - \frac12\bigl( 1 +\mu^2\, k\,\partial_k\bigr) P_{df}(k,\eta')  
\Bigr) \biggr] \,,
\eea
with
\be
M^A = \left(\begin{array}{cc}
0 & \frac12\\
0 & 0
\end{array}\right) \, , \qquad
M^B = \left(\begin{array}{cc}
\frac12 & 0\\
0 & \mu^2
\end{array}\right) \, .
\ee
Performing the angular average and taking the linear approximation for the long modes, i.e. $P_{ab}(q,\eta) \simeq u_a u_b \, P_L(q,\eta)$ with $u_a=(1,1)$, we obtain
\bea
\label{eq:squeezedB1}
\hskip -0.9 cm B_{abc}\!\left(\boldsymbol{k},-\boldsymbol{q},
\boldsymbol{q}-\boldsymbol{k},\eta\right)^{\textnormal{av}} 
&\!\xrightarrow{q \to 0}& 
\!u_b \, P_L\!\left(q,\eta\right)
\int_{\eta_0}^{\eta}\!
d\eta' \, g_{ad}\!\left(\eta,\eta'\right) g_{cf}\!\left(\eta,
\eta'\right) \, e^{-(\eta-\eta')} \,   \nn \\
&& 
\!\times \biggl[ 
M^{\rm av}_{dg}\, P_{gf}(k,\eta') + P_{dg}(k,\eta')\,  M^{\rm av}_{gf} - 
\Bigl(1 + \frac13 k\,\partial_k \Bigr)P_{df}(k,\eta')
\biggr],
\eea
and
\be
M^{\rm av} = 2 \left[ M^A + M^B \right]^{\rm av}
= \left(\begin{array}{cc}
1 & 1\\
0 & \frac23\\
\end{array}\right) \, ,
\ee
provided $Q_{abcd} = 0$. We would like to emphasize that up to this point we neither specified the background cosmology nor did we perform any perturbative expansion. 

This expression readily reproduces the linear results. For example, using an EdS background and choosing $\eta_{0}=0$, we get at
leading order\footnote{The linearized form of the expression in \eqref{eq:squeezedB1} also allows for different background cosmologies.} 
\beq
B^L_{abc}\!\left(\boldsymbol{k},-\boldsymbol{q},
\boldsymbol{q}-\boldsymbol{k},\eta\right)^{\textnormal{av}} 
\xrightarrow{q \to
0} 
 \, 
P_{L}\!\left(q,\eta\right)\,u_b  \biggl(
\frac1{21}
\left(\begin{array}{cc}
47 & 39\\
39 & 31\\ 
\end{array}\right)_{ac} 
- \frac13 u_a u_c \, k~\partial_k
\biggr)P_L\!\left(k,\eta\right)   \label{eq:squeezedB2}\, ,
\eeq
which coincides with the known result~\cite{Sherwin} for the component $B^L_{111}$ quoted in (\ref{eq:squeezedBlinear}). Note, at the same time, \eqref{eq:squeezedB2} is a generalization of \cite{Sherwin} which includes the velocity field.\\

While the result above is a generalization of the expression found in the literature at leading order, our actual aim was 
an expression of soft-limit relations which are valid in the non-perturbative regime. The previous relations, however, relies on the closure approximation. Although at linear order this is guaranteed by factorization, there is in principle no reason to believe this is a reasonable approach at higher orders or, more ambitiously, at non-perturbative level.  In fact, the trispectrum obeys an evolution equation that cannot be consistently set to zero at all times, despite what is often done in the literature, e.g. \cite{Pietroni:2008jx}. Therefore, it is important to assess to what extent the result (\ref{eq:squeezedB}) contains non-linear information. Using our results in section \ref{sec:SPT}, we can already check these relations at NLO. For that purpose, we insert the expression for the power spectrum at one-loop order in \eqref{eq:SPT_P_1loop} into the right hand side of (\ref{eq:squeezedB1}) for the short modes. Consequently, we find that the time-flow approach leads to 
\be
\alpha^{\rm TF} = \frac{103}{6930} \simeq 0.015   \, ,\quad
\beta^{\rm TF} = -\frac{233}{1890} \simeq -0.123 \, ,\quad
\gamma^{\rm TF} = \frac{271}{19404} \simeq 0.014 \, .\quad \label{4.4}
\ee
These numbers differ significantly from the SPT results in \eqref{4.2}. Hence, we conclude that the trispectrum cannot be ignored when computing the bispectrum beyond linear perturbation theory, which was the sole assumption on the way to derive \eqref{eq:squeezedB1}.

\section{Correlation Functions at Equal Times II: Locally Curved Universe\label{sec:5}}

We now move to an alternative approach to set up equal-time relations in the soft limit based on ideas introduced in \cite{baldauf2,Sherwin}, as well as the proposal for (angular averaged) `equal-time consistency conditions' of \cite{Valageas:2013zda} and \cite{Kehagias:2013paa}. We follow closely the analysis in \cite{baldauf2, Sherwin}, to which we urge the reader to consult for further details. At the same time, we present a shortened derivation of the relevant transformation(s) that avoids using the Fermi coordinates and directly matches a flat coordinate system with a soft mode to a (locally) curved one.

\subsection{Newtonian mapping}

We start from a FRW cosmology in global coordinates that includes a long-wavelength (soft) perturbation, $\Phi_L$. In the Newtonian approximation, we then have
\beq
ds^2=-\bigl[ 1+ 2\Phi_L(\bx,t)\,\bigr] dt^2 + a^2(t)\bigl[1 - 2\Phi_L(\bx,t)\bigr] d\bx^2\,.
\eeq 
Assuming that the perturbation is spherically symmetric, the Newton potential is related to the density contrast by Poisson's equation,\footnote{As it was shown in \cite{Creminelli:2013mca,Creminelli:2013poa}, the constant and gradient pieces of the potential can be removed by a change of coordinates to a free-falling frame. Such transformation also leads to the consistency conditions we re-derived previously in section \ref{sec:2}. Since we are interested in the {\it physical} squeezed limit, in this section we will only deal with the quadratic part of the potential, $\Phi_L \propto {\bx^2}$, and the equivalence with a locally curved universe discussed in \cite{baldauf2}.}  
\be
\label{eq:Poissonreal}
\Phi_L(\bx,t) \simeq \frac14 \, H^2 \, a^2 \delta_L \,\bx^2 \, \ll 1 \, .
\ee
We will search for a coordinate transformation of the form
\bea
t &=& t_K + f(t_K, \bx_K) \, , \nn \\
\bx &=& \bx_K (1 + g(t_K, \bx_K)) \, ,
\label{eq: trafo} 
\eea
with coordinate dependent functions $f$ and $g$, such that one can transform to a locally curved system where the metric (in isotropic coordinates)
takes the form
\beq
\label{ds2}
ds^2 = - d t_K^2 + a_K^2(t_K) \frac{d\bx_K^2}{\left(1 + \frac{1}{4} K \bx_K^2\right)^2} \, ,
\eeq
and $K$ correspond to the curvature. The time-time component of the metric enforces
\be
\dot f = - \Phi_L \, , 
\ee
while the vanishing time-space part yields
\be
a^2 \, \dot  g  \, \bx = \nabla f \, .
\ee
Furthermore, the scale factor also transforms\footnote{Notice that we are dealing with a subclass of possible perturbations, i.e. soft and spherically symmetric. Our ansatz is valid provided $\delta_L \ll 1$ and $ H^2 \bx^2 \ll 1$, such that $f\simeq \bx_K^2$ and $g\simeq \bx_K^0$. Otherwise, spatial gradients of $\delta_L$ are involved and the analysis has to be modified. However, these type of perturbations are sufficient for our purposes to study the squeezed limit of correlators.}
\be
a(t) = a(t_K)\,( 1 + H \, f) \, , 
\ee
and contributes to the curvature in the new coordinate system. In total, we find
\bea
a_K &=& a \,  (1 + g) \, , \nn \\
K  \, \bx^2 &=&   4 \Phi_L +  4 H \, \int dt  \, \Phi_L  \, .
\eea
In particular, the Hubble parameters in the two systems are related by
\be
H_K \equiv \frac{\dot a_K}{a_K} \simeq H + \dot g = H - \frac1{2a^2} \int dt \, a^2 \, H^2 \delta_L \, .
\ee
In EdS, the integrand is approximately constant in time (since $H^2 \propto a^{-3}$, $\delta_L \propto a$) and we get 
\bea
H_K &\simeq& H \left(1 - \frac13 \delta_L \right) \, , \label{eq: H_K} \\
a_K &\simeq& a \left(1 - \frac13 \delta_L \right) \, , \label{eq: a_K} \\
K &\simeq& \frac53 H^2 a^2 \delta_L\, \label{510}.
\eea
Note that these results are consistent with the Friedman equations in the curved coordinate system,
\be
H_K^2 = H^2 \frac{a^3}{a_K^3} - \frac{K}{a_K^2} \, .
\ee
Moreover, while the energy density in the curved system is larger
\be
\bar \rho_K = \bar \rho \,(1 + \delta_L) = \bar \rho  \frac{a^3}{a_K^3} \, ,
\ee
the physical Hubble rate $H_K$ in (\ref{eq: H_K}) is not, since the curvature over-compensates the density increase.

\subsection{The non-perturbative (physical) soft limit} 

We now turn on the short scale modes to study the imprint of a long-wavelength perturbation in their dynamics. These will be affected by the coordinate transformation through the change in expansion rate (\ref{eq: a_K}) and the additional `contraction' $(1+g)$ of the spatial coordinates (see (\ref{eq: trafo})). In particular, the density contrast of a hard mode transforms in real space as 
\be
\label{eq:trafo_delta}
\delta_K(\bx_K, a_K) = \delta(\bx, a)\, (1 - \delta_L) \, ,
\ee
so that the corresponding two-point correlation function, defined as $\xi(r)\equiv \langle \delta(\bx)\,\delta(\bx+\br) \rangle$,  reads
\bea
\label{eq:xidelta}
\xi_{\delta_L}(r,a) 
&=&  \left[1+ \delta_L \left( 2 + 
\frac{1}{3} r \partial_r  
\right) \right]\xi_K(r,a_K) \nn \\
&=&  \left[1+ \delta_L \left( 2 + 
\frac{1}{3} r \partial_r  
-\frac13 \partial_\eta
\right) \right]\xi_K(r,a) \, .
\eea
Here, $\xi_K(r,\eta)$ is the correlation function in the curved coordinate system, and we used again $r_K = r\, ( 1- g)$, $a_K = a\, ( 1+ g)$ together with the EdS result $g = -\delta_L/3$.

The local curvature also affects the growth of structure. In the case of an EdS universe, for instance, the linear growing mode 
of the density contrast becomes 
\bea
D(K, a_K) &\simeq&
D(0, a)  
+   \left. K \frac{d}{dK} D(K,a_K) \right|_{K=0}\nn \\
&=& \left( 1+ \frac{13}{21} \delta_L\right) D(0,a)\,. \label{eq:growth}
\eea
This means at linear order we may replace
\beq
\label{eq:xi0}
\xi^{(1)}_K(r,a_K) \to \left(1+ 2\times \frac{13}{21} \delta_L\right) \xi^{(1)}(r,a)\,,
\eeq
where $\xi(r,\eta)$ represents the correlation function in the original (flat) frame without the soft mode. \\

The combination of all the factors in \eqref{eq:xidelta} and \eqref{eq:xi0}, when correlated with $\delta_L$, gives the relation in the soft limit of the bispectrum at leading order, which was originally derived by Sherwin and Zaldarriaga in \cite{Sherwin} and coincides with \eqref{eq:squeezedBlinear} (see also \eqref{eq:squeezedB2}). However, as advocated in \cite{baldauf2}, absorbing the soft mode into a curved background is a correct procedure also at the non-perturbative level. For example, this has led to a physical squeezed limit for correlation functions in the inflationary case \cite{Creminelli:2013cga}. Following similar steps as in \cite{Creminelli:2013cga}, one can then write for the density perturbations in an EdS cosmology, 
\bea
B_{111}\!\left(\boldsymbol{k},-\boldsymbol{q},
\boldsymbol{q}-\boldsymbol{k},\eta \right)^{\rm av} 
&\xrightarrow{q \to 0}& 
P_L (q,\eta) \left[\left( 1 - \frac{1}{3}k~\partial_k -\frac13 \partial_\eta\right) P(k,\eta)
+ \frac{5}{3} \left.\frac{\partial}{\partial\kappa} P_K(k,\eta)\right|_{K=0}\right]  \, , \nn \\
\label{eq:BsoftK}
\eea
where $\kappa=K/(a^2H^2)$ and $P_{K}(k,\eta)$ is the power spectrum for the density contrast in the presence of (local) curvature. This expression follows simply from (\ref{eq:xidelta}), using the (linear) map between $K$ and $\delta_L$  given by \eqref{510}.

The relation in \eqref{eq:BsoftK}, although generic, is not the type one would expect to confront against observations,
since it involves a correlation function in a hypothetical case of a curved universe. Moreover, the impact of curvature on the fluctuations, i.e. $\partial P_K(k,\eta)/\partial K$, cannot be readily obtained in terms of quantities at $K=0$ without resorting to perturbation theory. The work of \cite{Kehagias:2013paa, Valageas:2013zda}, on the other hand, is an attempt at precisely achieving this. We will study their proposal next. 

\subsection{The `equal-time consistency relations'}\label{sec:41}

A proposal by Valageas, and also Kehagias, Perrier and Riotto (VKPR) to extend the computation in \cite{Sherwin} into the non-linear regime appeared in \cite{Valageas:2013zda} and \cite{Kehagias:2013paa} , where it was coined the (angular averaged) `equal-time consistency conditions' for large-scale structure. In practice, it consists of replacing each growth function for the short modes with a factor of \eqref{eq:growth}, yielding
\beq
\label{eq:xikprv}
\xi^{\rm VKPR}_{\delta_L}({r},\eta) = \left[1+\delta_L \left( 2 + \frac{1}{3} r \partial_r  + \frac{13}{21} \partial_\eta \right)\right] \xi(r,\eta)\,,
\eeq
or in Fourier space, after correlating with a long-wavelength mode, 
\be
\label{eq:kprv}
B_{111}^{\rm VKPR}\!\left(\boldsymbol{k},-\boldsymbol{q},
\boldsymbol{q}-\boldsymbol{k},\eta\right)^{\rm av} \xrightarrow{q \to
0} P_L (q,\eta) \left[ 2- \frac{1}{3}\left(3 + k~\partial_k \right)+ \frac{13}{21} \partial_\eta \right] P(k,\eta) \,, 
\ee
with the shortened notation $P(k,\eta) \equiv P_{11}(k,\eta)$ for the density power spectrum. (Recall $\eta \equiv \ln a(\tau)$, and therefore the logarithmic derivative serves as a counter.) 

As a first step, we can check the validity of \eqref{eq:kprv} at NLO by inserting the one-loop power spectrum (see \eqref{eq:SPT_P_1loop}) on the right hand side of \eqref{eq:kprv}. For the coefficients defined in section \ref{sec:SPT} we consequently find
\be
\alpha^{\rm VKPR}  = \frac{61}{1890}\simeq 0.032 \, ,\quad
\beta^{\rm VKPR} = -\frac{3599}{13230} \simeq -0.272 \, ,\quad
\gamma^{\rm VKPR} = \frac{135}{1372} \simeq 0.098 \, ,\quad
\ee
which overall differ from the SPT results. We notice, nonetheless, that $\alpha^{\rm VKPR} = \alpha^{\rm SPT}$. This must be indeed the case since it comes from the eikonal phase in \eqref{eq:factor}, which we argued  is universal. 

It is clear from our previous analysis that the subtle step in the derivation is the replacement    
\beq 
\label{eq:xin}
\xi^{\rm VKPR}_K(r,a_K) = \left(1+ \frac{13}{21} \delta_L \,\partial_\eta\right)\xi (r,a) \,,
\eeq
that aims at generalizing \eqref{eq:xi0}. As we see in the comparison, this replacement is not valid at higher orders even in an EdS background, since it fails to capture relevant dynamics.

We can already start to see the seed of the problem in the perturbative expansion for the density field itself before computing the correlation with the long mode. As discussed in \cite{baldauf2}, a simple exercise consists of using SPT to compute the density perturbation to a given order and check whether the above replacement correctly accounts for the effect of the long mode.  

This was performed directly in \cite{Sherwin} up to second order, i.e. $\delta^{(2)}(\bx,\eta)$, where one finds 
\beq
\label{eq:SZ}
\delta_{S,\delta_L}(\bx,\eta) =  \delta^{(1)}_{S}(\bx + \bd(\bx,\eta),\eta) + \frac{34}{21} \delta^{(1)}_L(\bx,\eta)\, \delta_S^{(1)} (\bx,\eta)+ \frac{4}{7} K_{ij}^{L}(\bx,\eta)\, K_{ij}^{S}(\bx,\eta) + \cdots~,
\eeq
where the ellipses represent higher-order terms in SPT. The displacement and anisotropy terms are defined as 
\bea
\bd(\bx,\eta) &\equiv& - \int_{\bq} \frac{i \bq}{q^2} \,\delta^{(1)}(\bq,\eta)\, e^{i\bq\cdot\bx}\,, \quad 
K_{ij}(\bx,\eta) \equiv \int_{\bq} \left(\frac{\bq_i \bq_j}{q^2}-\frac{1}{3} \delta_{ij}\right) \delta^{(1)}(\bq,\eta) \, e^{i\bq\cdot\bx} \,,
\eea
with $\int_{\bf q} \equiv \int \frac{d^3 \bq}{(2\pi)^3}$. Note that we have resummed the effects of the displacement to all orders, which follows from the eikonal approximation in \eqref{eq:factor}, or can be also shown directly in Lagrangian space \cite{left}. 
From here it is straightforward to derive the soft limit of the bispectrum by computing  
\beq \langle \xi_{\delta_L}(r,\eta)\,\delta_L\rangle = \langle \delta_{S,\delta_L}(\bx+\br,\eta)\, 
\delta_{S,\delta_L}(\bx,\eta) \,\delta_L \rangle\,,
\eeq 
and averaging over angles. The piece from the displacement reproduces the factor of $\tfrac{1}{3} r \,\partial_r \xi(r)$ in \eqref{eq:xikprv}, while the $\left(1 + \frac{13}{21}\right)$ (twice) gives the correct factor of $\frac{68}{21}$ found in \cite{Sherwin}. It was crucial in this derivation that the anisotropy term did not contribute upon angular averaging, $\langle \delta_L(\bx,\eta)\, K_{ij}^L(\bx,\eta) \rangle^{\rm av} \to 0 \,$.
This is in fact what follows for spherically symmetric density perturbations and is accomplished by the angular integral.

At NLO, the validity of the proposal in \cite{Kehagias:2013paa,Valageas:2013zda} requires that the density perturbation may be written as 
\beq
\label{eq:SZ_NLO}
\delta_{S,\delta_L}(\bx,\eta)  \supset  \delta^{(2)}_{S}(\bx + \bd(\bx,\eta),\eta) + \frac{47}{21} \delta^{(1)}_L(\bx,\eta) \,\delta_S^{(2)} (\bx,\eta)+ \frac{4}{7} K_{ij}^{L}(\bx,\eta)\, A_{ij}^{S}(\bx,\eta) + \cdots \,,
\eeq
where the ellipses would include higher-order terms. The factor $47/21$ would be a consequence of $\delta^{(2)} \propto D(\eta)^2$, as dictated by the expression in \cite{Kehagias:2013paa,Valageas:2013zda}. We also collected the anisotropy piece into the term containing the matrix $A^S_{ij}$, which vanishes once the soft mode is averaged over angles. However, as expected from the one-loop check of the explicit bispectrum calculation, this relation is not fulfilled. In fact, after averaging, we find an additional piece (see \eqref{533}) that leads to the (small) discrepancy between  the $\beta, \gamma$ coefficients computed in SPT and the proposal of \cite{Kehagias:2013paa,Valageas:2013zda}.

Albeit formally not valid at one-loop order, the errors in the coefficients $\beta$ and $\gamma$ are small. For instance including up to fourth order, i.e. $\delta^{(4)}(\bx,\eta)$, the calculation shows that the claim  in  \cite{Kehagias:2013paa,Valageas:2013zda} for the form of the density perturbation is quantitatively (very) close, but not quite the same as the one we expect from the SPT result.\footnote{Notice that although the errors in the coefficients $\beta$ and $\gamma$  are small, they are still multiplied by an integral which, as we emphasized, needs to be regularized. This means that the discrepancy may be ultimately large (or even diverge) for initial conditions where the integrals are dominated by the hard part of the spectrum.} We will elaborate on the reasons behind the discrepancy in section \ref{sec:discp}. However, we can already start by identifying plausible causes, in particular for the case of the velocity field as we discuss next.\footnote{Regarding the assumption of spherical symmetry, one may be worried that anisotropy terms could survive after angular averaging. Even though we are taking the soft limit $q\to 0$, the angular dependence coming from the soft mode remains because of the $1/q^2$ enhancement from the eikonal phase, e.g. $\frac{(\bk\cdot \bq)^2}{q^2}$. However, due to factorization, at equal times the anisotropy terms vanish upon averaging. Or in other words, we do not encounter singularities of the form $1/q^4$ at equal times, since the contributions from the eikonal phase cancel each other and the remaining terms are analytic in $q$. This can be explicitly checked up to NNLO.} 

\subsection{The impact of the velocity}

To be precise, the relation (\ref{eq:growth}) only holds for the growing mode of the density contrast, but not for the velocity. As we discuss next, the response of the velocity fields to the presence of a long-wavelength mode is different.
 
The dependence of the growing mode on curvature comes from two sources 
\be
\frac{d D(K,a_K)}{d\delta_L} = 
\frac{\partial D}{\partial K} \frac{dK}{d\delta_L} +
\frac{\partial D}{\partial a_K} \frac{da_K}{d\delta_L} \, .
\ee
While the first term contributes $20/21$ and describes how the growing mode as a function of $a$ (or $\eta$) is modified by curvature, the second term gives $-1/3$ and accounts for the change in the scale factor when the growing mode is compared at the same proper time. In total, the growing mode of the density contrast increases by a factor $(1 + 13/21 \delta_L)$ in the 
locally curved system \cite{baldauf2,Sherwin}. Transforming back using (\ref{eq:trafo_delta}) yields that the density growth is enhanced by a factor $(1 + 34/21 \delta_L)$ in presence of a soft mode. 

Repeating the same calculation for the velocity field, one finds that the growing mode of $\Theta$, defined in (\ref{eq: fields}), responds differently to curvature. In particular, $\partial D_{\Theta}/\partial K$ is twice as large. (Note that this is also required by the continuity equation.)  Hence, the velocity field grows faster by a factor $(1 + 33/21 \delta_L)$ in the curved universe.  In order to obtain the final expression for the change in $\Theta$ we also need to include the effect of the coordinate transformation. First, we note that the physical velocity $\boldsymbol{v}$ is the same in the two systems, and therefore $\Theta$ transforms as 
\be
\Theta_K (\br_K, a_K) = \Theta (\br, a)\, (1+ \delta_L/3) \, . 
\ee
In total, the velocity then grows faster by a factor $(1 + 26/21 \delta_L)$ in presence of a soft mode,  which is consistent
with the explicit SPT calculation (for example~$2 \times 26/21 - 1 = 31/21$, c.f.~(\ref{eq:squeezedB2})). 

This result has important consequences. First of all, it means that a relation similar to \eqref{eq:kprv} which includes the velocities may be (naively) written down through an extension of the linear result (including the other components of \eqref{eq:squeezedB2}) by use of the above transformations. Following the steps in \cite{Kehagias:2013paa,Valageas:2013zda} translated to the velocity one would then write 
\beq
\label{eq:SZ_NLOth}
\Theta_{S,\delta_L}(\bx,\eta)  \supset  \Theta^{(n)}_{S}(\bx + \bd(\bx,\eta),\eta) +\delta^{(1)}_L(\bx,\eta)  \left(-\frac{1}{3}  + \frac{33}{21} \partial_\eta\right) \Theta_S^{(n)} (\bx,\eta)+ \cdots \,.
\eeq
Here the ellipses would represent terms which either vanish when an angular average is performed or are higher order.\footnote{Notice we could also replace $\delta^{(1)}_L \to \Theta^{(1)}_L$.} However, one can show that the expression in \eqref{eq:SZ_NLOth} dramatically fails beyond linear order (see below). This is already a signal that one has to carefully account for the impact of the velocities in the SPT computations. 

\subsection{Fluid perturbations in a curved background}\label{sec:discp}

In the standard SPT manipulations, one replaces $\delta^{(1)}$ by $\Theta^{(1)}$.  However, as we just showed, the velocity and density fields respond differently to the presence of a long-wavelength perturbation. Unfortunately, there is no easy way to deduce the dependence of the non-linear power spectrum on the two growing and decaying modes of $\psi$, separately. This would be necessary to account for the different curvature effects in each component. In principle, one would then expect a significant departure from the relation in \eqref{eq:kprv} beyond linear order. In practice, on the other hand, what the above computations unravel (at least perturbatively) is that, for the case of the density field, the relation in \eqref{eq:kprv} qualitatively reproduces the SPT result. In order to gain some intuition behind this small discrepancy, we will next inspect in more detail the dynamics of fluid fluctuations in a curved background in perturbation theory.

Consider the fluid equations in (\ref{eq:  fluid equations}) for the short modes in a curved background. Adding curvature to the cosmology modifies the matrix $\Omega_{ab}$, and at leading order in $K \propto \delta_L$ the corresponding contribution for an EdS cosmology reads 
\begin{equation}
\label{eq:OmegaK}
\Omega_{K,ab} \simeq
\Omega_{K=0,ab} + 
\left. K \frac{\partial}{\partial K}\Omega_{K,ab} \right|_{K=0} =
\Omega_{ab} +  \kappa \left(\begin{array}{cc}
\;\;\,0 & \quad \;\;0\\
-3/2 & \;\;\,-1/2
\end{array}\right) \, .
\end{equation}
The additional piece will modify the propagator in a fully non-perturbative treatment. However, since the long-wavelength mode may be treated within perturbation theory, at leading order in $\delta_L$ it suffices to treat the effect of curvature as an extra interaction. Note that this extra term is time dependent since $ \kappa \propto a(\eta)$. (The curvature, $K$, is on the other hand time independent.)  

At linear order in the fluctuations for the short modes we consequently have
\begin{equation}
\begin{alignedat}{1}\psi_{K}^{(1)}(\bk_{1},\eta) & \simeq\psi_{K=0}^{(1)}(\bk_{1},\eta)+K\left.\frac{\partial}{\partial K}\psi_{K}^{(1)}(\bk_{1},\eta)\right|_{K=0}\\
 & =\left[\left(\begin{array}{c}
1\\
1
\end{array}\right)+\,\frac{4\kappa}{7}\left(\begin{array}{c}
1\\
2
\end{array}\right)\right]e^{\eta-\eta_{0}}\,\delta^{(1)}(\bk_{1},\eta_{0})\,.
\end{alignedat}
\end{equation}

Note, as we mentioned before, the impact of curvature on the velocity is twice as large as the response of the density field. 

In general there are two contributions to the $n$-th order solution, $\psi_K^{(n)}$, at linear order in~$K$. The first arises from the additional interaction (just as the leading-order term)
\be
\label{eq:curvCont1}
K \int \, d\eta' \, g_{ab}(\eta,\eta') \, \frac{\partial}{\partial K}\Omega_{bc}(\eta') \, \, \psi_c^{(n)}(\bk_1, \eta') \, ,
\ee
while the second contribution stems from the $K$-dependence in the lower order solutions that 
enter in the SPT recursion relation
\be
\label{eq:curvCont2}
K \int \, d\eta' \, g_{ab}(\eta,\eta') \, \gamma_{bcd}(\bk,-\bk_1,-\bk_2) \, 
\frac{\partial}{\partial K} \left[ \psi_{K,c}^{(n-m)}\!(\bk_1, \eta') \, \psi_{K,d}^{(m)}(\bk_2, \eta') \right] \, .
\ee
It is important to stress that, already at NLO, $\partial \psi_K^{(2)}(\bk,\eta) / \partial K$ is not {\it proportionally} related to $\psi_{K=0}^{(2)}(\bk_1,\bk_2)$. 
Nonetheless, we find the following relation,
\bea
\label{533}
K \left.\frac{\partial}{\partial K} \psi_K^{(2)}(\bk,\eta)\right|_{K=0} &=& 
\, \frac{4 \kappa}{7}
\left(\begin{array}{cc}
2 & 0\\
0 & 3
\end{array}\right) \psi_{K=0}^{(2)}(\bk,\eta) \nn \\
&& \, + \, 
\frac{\kappa}{147} \left(\begin{array}{c} 1\\3 \end{array}\right)
\int \, d\bk_1 \, d\bk_2 \, \delta^3(\bk+\bk_1+\bk_2) \\
&& \quad \times 
\frac{(\bk_1 \cdot \bk_2)^2 - k_1^2 k_2^2}{k_1^2 k_2^2}  
\, e^{2(\eta-\eta_0)} \, \delta^{(1)}(\bk_1,\eta_0)  \, \delta^{(1)}(\bk_2,\eta_0) \, \nn . 
\eea
Hence, the overall difference with the expression in \eqref{eq:kprv} for the density contrast (first entry) is rather small. At the same time, we see that the extension for the velocities in \eqref{eq:SZ_NLOth} fails (compare with second entry).

Even though we conclude that the curvature dependence cannot be naturally reformulated in terms of a derivative with respect to $\eta$, as suggested by the proposal in \cite{Kehagias:2013paa, Valageas:2013zda}, we find that (extending the result in \eqref{533}) 
\be
\label{535}
K \left.\frac{\partial}{\partial K} \psi^{(n)}_K(\bk_{K},\eta)\right|_{K=0} \simeq
\, \frac{4 \kappa}{7}
\left(\begin{array}{cc}
\partial_{\eta} & 0\\
0 & \,\partial_{\eta}+1
\end{array}\right)
\psi^{(n)}_{K=0}(\bk_{K},\eta) \, ,
\ee
gives a reasonable {\it empirical} perturbative approximation, which translates into
\be
\label{536}
\left.\frac{\partial}{\partial \delta_L}\psi^{(n)}_{\delta_L}(\bk,\eta)\right|_{\delta_L=0} \simeq 
\, \frac{1}{21}
\left(\begin{array}{cc}
21+13\,\partial_{\eta} & 0\\
0 & \,13+13\,\partial_{\eta}
\end{array}\right) 
\psi^{(n)}_{K=0}(\bk_{K},\eta) \, .
\ee

This expression may be motivated as follows. Starting from the fluid equations in \eqref{eq:  fluid equations}, redefining the velocity as $\Theta \to \Theta/f$ and making the change of time variable  $\eta \to \log D_K(\eta)$, $D_K(\eta)$ being the growth factor in the presence of curvature, we get a new set of fluid equations with\footnote{We used $\Omega_m \simeq 1+ \kappa$ and $f  \simeq 1 + \frac{4}{7} \kappa$, to leading order in $\kappa$ for an EdS cosmology in the presence of curvature.}
\begin{equation}
\label{eq:Omegaf}
\Omega_{f,ab} =
\left(\begin{array}{cc}
\;\;\,0 & -1\\
-\frac{3}{2}\frac{\Omega_m}{f^2} & \;\;\, \frac{3}{2}\frac{\Omega_m}{f^2} -1
\end{array}\right)
\simeq \Omega_{ab} +  \frac{3\kappa}{14} \left(\begin{array}{cc}
0 & \;\;\;\,0\\
1 & \;-1
\end{array}\right) \ ,
\end{equation}
and
\be 
f\equiv \frac{\partial \log D_K}{\partial \log a} \, .
\ee
The relation in (\ref{535}) is obtained if one neglects the term proportional to $\kappa$ in (\ref{eq:Omegaf}). In other words, the dependence on the cosmology has been absorbed into the growth factor in the presence of curvature, hence derivatives with respect to $K$ may be traded by derivatives with respect to $\eta$. Notice that the new contribution in $\Omega_{f,ab}$ annihilates the growing mode, i.e.  
$\psi^{(n)} \propto (1,1)$, which in turn explains the relative accuracy of (\ref{535}) in the SPT computations.  

\section{Discussion}\label{sec:disc}

In this work, we studied correlators of the density and velocity fields in the soft limit. We first (re-)derived the well known consistency conditions at unequal times using the eikonal approximation, which naturally accounts for the resummation of the soft mode. This provides a compact expression that also generalizes to include the velocity field. 

Afterwards, we explored under which circumstances equal-time relations exist. For that purpose we computed the bispectrum beyond leading order in SPT to assess two approaches that aim at deriving (allegedly non-perturbative) expressions. The first is based on the time-flow approach in the Eulerian representation. In this scheme, the connected four-point function, namely the trispectrum, is neglected with questionable success in computations involving the power spectrum \cite{Pietroni:2008jx, Audren:2011ne}. An attempt at including the information from the trispectrum appeared in \cite{Juergens:2012ap}, which points towards a non-negligible contribution already in the mildly non-linear regime. In principle, one could have hoped that the trispectrum was less relevant in the soft limit, and equal-time relations from the time-flow approach may be approximately accurate. Unfortunately, this is not the case and we found large deviations from SPT already at one-loop order for the bispectrum. Including the trispectrum and truncating the hierarchy at higher orders would reproduce the one-loop SPT result, however, it will fail at some given loop order depending on the truncation. Therefore, a truly non-perturbative result seems out of reach in the time-flow formalism.

Overall, in spite of not holding up to the same status as the unequal-time relations, perturbative statements between correlation functions at equal times in the squeezed limit may be still useful in special circumstances in which the short modes may be kept in the mildly non-linear regime. The perturbative relations of the time-flow approach are well suited, for instance, to study baryonic acoustic oscillations in a background cosmology that requires numerical input, e.g.~models including massive neutrinos or quintessence. Unlike SPT, the time-flow approach only deals with equal-time quantities, such that soft effects cancel out from the outset. These relations may thus improve numerical stability and aid the computational treatment of the fluctuations. \\

We then moved to the curved background method and the (angular averaged) `equal-time consistency conditions' of Valageas, and also Kehagias, Perrier and Riotto \cite{Kehagias:2013paa,Valageas:2013zda}. Their proposal was an attempt to use an equivalence discussed in \cite{baldauf2}, between physics on short scales in the presence of a long-wavelength perturbation and a locally curved cosmology, to extend the leading result in \cite{Sherwin} into the non-linear regime. In \cite{baldauf2} it was argued that the physical equivalence applies even when the short modes are deep in the non-linear regime. As we discussed, this equivalence relates, for example, the bispectrum of density fluctuations in the soft limit to the variation of the power spectrum on short scales in the presence of local curvature $K$ in an EdS cosmology,
\bea
B \!\left(\boldsymbol{k},-\boldsymbol{q},
\boldsymbol{q}-\boldsymbol{k},\eta \right)^{\rm av} 
&\xrightarrow{q \to 0}& 
P_L (q,\eta) \left[ \left(1 - \frac{1}{3}k~\partial_k -\frac13 \partial_\eta \right) P (k,\eta) 
+ \frac{5}{3} \left.\frac{\partial}{\partial \kappa}  P_K(k,\eta)\right|_{K=0}\right]  \, , \nn \\
\label{eq:BsoftKdisc}
\eea
with $\kappa=K/(a^2H^2)$ (see \eqref{eq:BsoftK}). The relation (\ref{eq:BsoftKdisc}) can be shown to be universal, hence valid non-perturbatively. The term involving $\left(1-\tfrac{1}{3} k\partial_k\right)$ is a combination of two effects, namely, the difference in the density contrast between the two cosmologies plus the shift induced by the displacement term \cite{left} (or the eikonal phase); while the contribution $\tfrac{1}{3}\partial_\eta$ follows from the change in the scale factor. It is not, however, an expression that can be directly confronted with observations, 
since it involves the dependence of the power spectrum on a hypothetically (locally) curved universe through the last term. One may, nevertheless, use numerical simulations to determine the derivative with respect to $\kappa$. 
This was performed in \cite{Li:2014sga, Wagner:2014aka} in the so-called `separate universe' approach.
The proposal in \cite{Kehagias:2013paa, Valageas:2013zda}, on the other hand, can be rephrased as an attempt to replace the variation with respect to curvature with quantities which can be computed when $K=0$, more specifically
\beq
\label{disc1}
{\rm VKPR:} \quad 
 \left. \frac{\partial}{\partial \kappa} P_K(k,\eta)\right|_{K=0} = 
\frac{4}{7}  \, \partial_\eta P_{K=0}(k,\eta)\,.
 \eeq

As we showed, re-writing the fluid equations in terms of a new time variable using the growth factor in the presence of curvature, i.e. $\eta \to \log D_K(\eta)$, one can absorb the information regarding the background cosmology into the time evolution, up to a $\kappa$-dependent interaction (see \eqref{eq:Omegaf}). Neglecting this (time-dependent) extra term implies the relation in \eqref{disc1} for the density contrast (and more generally the expression in \eqref{535} including the velocity field.) Assessing the accuracy of \eqref{disc1} thus amounts to estimating the error induced in this approximation. While we have explicitly demonstrated that the expression (\ref{disc1}) does not fully account for the effect of local curvature on the growth of structure, we have also found that \eqref{disc1} is quantitatively accurate for the power spectrum of density fluctuations in the realm of perturbation theory, to the few-percent level. This was confirmed by a one-loop check of the bispectrum, and also by explicitly computing the density fluctuations in SPT (see \eqref{535}.)

At first, the accuracy of the proposal in \cite{Kehagias:2013paa, Valageas:2013zda} may be related to the variation  \beq \frac{\partial}{\partial{\delta_L}}\left(\frac{3}{2}\frac{\Omega_m}{f^2}\right) \simeq - \frac{5}{14},\eeq
in the presence of a long-wavelength fluctuation that has been absorbed into the background. Notice, however, that the precision we find in perturbative computations is much better, and it is in fact due to extra cancellations. At leading order in $\kappa$, the additional term in the fluid equations \eqref{eq:Omegaf} almost annihilates the EdS solution at any given order in SPT, which is dominated by $\delta^{(n)} \simeq \Theta^{(n)}$ (i.e. $\psi^{(n)} \propto (1,1)$). 
While this explains the {\it unreasonable effectiveness} in perturbative computations, it does not necessarily imply a similar accuracy in the non-perturbative regime, which requires additional study.\footnote{As discussed in \cite{Valageas:2013zda}, exact relations can be found in a simplified (1+1 dimensional) toy model, whose background equations resemble an EdS cosmology. However, in this example it is easy to see that the response to a soft mode is given by $\psi_{\delta_L}\simeq [1 + \delta_L (1 + \partial_\eta)]\, \psi$; and moreover, the Zel'dovich approximation \cite{Zeldovich:1969sb} is exact with fluctuations always remaining in the growing mode. Hence, both density and velocity respond equally to curvature. Unfortunately this does not provide any additional insight for the approximate validity of \eqref{disc1}. Attempts at testing the proposal of  \cite{Kehagias:2013paa, Valageas:2013zda} against numerical simulations appeared in \cite{Chiang:2014oga,checkkprv}. The small deviations found in both cases are consistent with our findings.}
\begin{center}
\bf{Acknowledgements}
\end{center}
We thank Mathias Garny, Daniel Green and Matias Zaldarriaga for helpful comments. This work was supported in part by the German Science Foundation (DFG) within the Collaborative Research Center (SFB) 676 `Particles, Strings and the Early Universe.' 

\appendix

\section{The Bispectrum at One-loop Order in SPT\label{appBispectrum}}

The NLO correction to the linear bispectrum
of density perturbations, i.e. the one-loop contribution $B_{111}^{\rm 1-loop}$, 
is given as the sum of four diagrams involving density correlations \cite{Bernardeau:2001qr},
\begin{equation}
B_{111}^{\rm 1-loop}\!\left(k_{1},k_{2},k_{3},\eta\right)=B^{222}+B_{I}^{321}+B_{II}^{321}+B^{411}\,.
\end{equation}
Taking the UV-limit for the loop momenta, $l\gg k_{1,2,3},$ these contributions take the form~\cite{Baldauf:2014qfa}
\begin{eqnarray}
B^{222}\!\left(k_{1},k_{2},k_{3},\eta\right) & \simeq & -\frac{1}{4802\pi^{2}}\Bigl[30k_{1}^{6}-30k_{1}^{4}\bigl(k_{2}^{2}+k_{3}^{2}\bigr)+k_{1}^{2}\bigl(-30k_{2}^{4}+k_{2}^{2}k_{3}^{2}-30k_{3}^{4}\bigr) \label{eq:B222} \\
 &  & +30\bigl(k_{2}^{2}-k_{3}^{2}\bigr)^{2}\bigl(k_{2}^{2}+k_{3}^{2}\bigr)\Bigr]\,\int\! dl\, l^{2}\left(\frac{P^{3}_{L}\!\left(l,\eta\right)}{l^{6}}\right)\nn\,, 
\end{eqnarray}
\begin{eqnarray}
B_{I}^{321}\!\left(k_{1},k_{2},k_{3},\eta\right) & \simeq & \frac{1}{35280\pi^{2}k_{3}^{2}}\Bigl[170k_{1}^{6}+k_{1}^{4}\bigl(83k_{2}^{2}+190k_{3}^{2}\bigr)+2k_{1}^{2}\bigl(67k_{2}^{4}+256k_{2}^{2}k_{3}^{2}  \\
 &  & -445k_{3}^{4}\bigr)-\bigl(387k_{2}^{2}-530k_{3}^{2}\bigr)\bigl(k_{2}^{2}-k_{3}^{2}\bigr)^{2}\Bigr]\, P_{L}\!\left(k_{3},\eta\right)\int\! dl\, l^{2}\left(\frac{P_{L}^{2}\!\left(l,\eta\right)}{l^{4}}\right)\nn  \\
 &  & +\textnormal{ 5 perm.}\nn\,,
\end{eqnarray}
\begin{eqnarray}
B_{II}^{321}\!\left(k_{1},k_{2},k_{3},\eta\right) & \simeq & -\frac{61}{105}\, F_{2}\!\left(\boldsymbol{k}_{2},\boldsymbol{k}_{3}\right)\, P_{L}\!\left(k_{2},\eta\right)P_{L}\!\left(k_{3},\eta\right)k_{3}^{2}\int\! dl\, l^{2}\left(\frac{P_{L}\!\left(l,\eta\right)}{l^{2}}\right)  \\
 &  & +\textnormal{ 5 perm.}\nn\,,
\end{eqnarray}
and 
\begin{eqnarray}
\label{eq:B411}
B^{411}\!\left(k_{1},k_{2},k_{3},\eta\right) & \simeq & -\frac{1}{226380}\frac{1}{k_{2}^{2}k_{3}^{2}}\Bigl[12409k_{1}^{6}+20085k_{1}^{4}\bigl(k_{2}^{2}+k_{3}^{2}\bigr) \\
 &  & +k_{1}^{2}\bigl(-44518k_{2}^{4}+76684k_{2}^{2}k_{3}^{2}-44518k_{3}^{4}\bigr)\nn \\
 &  & +12024\bigl(k_{2}^{2}-k_{3}^{2}\bigr)^{2}\bigl(k_{2}^{2}+k_{3}^{2}\bigr)\Bigr]\, P_{L}\!\left(k_{2},\eta\right)P_{L}\!\left(k_{3},\eta\right)\int\! dl\, l^{2}\left(\frac{P_{L}^{2}\!\left(l,\eta\right)}{l^{2}}\right)
\nonumber 
\\
 &  & +\textnormal{ 2 perm.}\nn
\end{eqnarray}
Thereby, $F_{2}\!\left(\boldsymbol{k}_{2},\boldsymbol{k}_{3}\right)$
denotes the symmetrized second-order kernel in SPT and the permutations
have to be taken with respect to the external momenta.\\

We now rewrite (\ref{eq:B222})-(\ref{eq:B411})
in terms of the momenta $\boldsymbol{k}$ and $\boldsymbol{q}$, average
over the respective angles and finally take the soft-$q$ limit. While
\begin{equation}
B^{222}\!\left(\boldsymbol{k},-\boldsymbol{q},\boldsymbol{q-k},\eta\right)^{\textnormal{av}}\xrightarrow{q\to0}\,0 \,,
\end{equation}
the other resulting expressions are given by
\begin{eqnarray}
B_{I}^{321}\!\left(\boldsymbol{k},-\boldsymbol{q},\boldsymbol{q-k},\eta\right)^{\textnormal{av}}\! & \xrightarrow{q\to0} & P_{L}\!\left(q,\eta\right) \frac{k^{4}}{\pi^{2}}\,\gamma^{\textnormal{SPT}}\!\int\! dl\, l^{2}\!\left(\frac{P_{L}^{2}\!\left(l,\eta\right)}{l^{4}}\right)\!,  \\
B_{II}^{321}\!\left(\boldsymbol{k},-\boldsymbol{q},\boldsymbol{q-k},\eta\right)^{\textnormal{av}}\! & \xrightarrow{q\to0} & P_{L}\!\left(q,\eta\right)\frac{k^{2}}{\pi^{2}}\!\left( \alpha_1^{\textnormal{SPT}}\, k\partial_{k}+\beta_{1}^{\textnormal{SPT}}\!\right)\! P_{L}\!\left(k,\eta\right)\!\int\! dl\, l^{2}\!\left(\frac{P_{L}\!\left(l,\eta\right)}{l^{2}}\right)\!,\\
B^{411}\!\left(\boldsymbol{k},-\boldsymbol{q},\boldsymbol{q-k},\eta\right)^{\textnormal{av}}\! & \xrightarrow{q\to0} & P_{L}\!\left(q,\eta\right)\frac{k^{2}}{\pi^{2}}\!\left( \alpha_2^{\textnormal{SPT}}\, k\partial_{k}+\beta_{2}^{\textnormal{SPT}}\!\right)\! P_{L}\!\left(k,\eta\right)\!\int\! dl\, l^{2}\!\left(\frac{P_{L}\!\left(l,\eta\right)}{l^{2}}\right)\!, 
\end{eqnarray}
with
\begin{equation}
\gamma^{\textnormal{SPT}}=\frac{515}{5292}\,,
\quad\alpha_1^{\textnormal{SPT}}=\alpha_2^{\textnormal{SPT}}=\frac{61}{3780}\,,
\quad\beta_{1}^{\textnormal{SPT}}=-\frac{671}{8820}\,,
\quad\beta_{2}^{\textnormal{SPT}}=-\frac{155}{756}\,.
\end{equation}
Finally, the sum of these contributions gives the total SPT one-loop correction to the bispectrum, yielding
\begin{eqnarray}
B_{111}^{\textnormal{1-loop}}\!\left(\boldsymbol{k},-\boldsymbol{q},\boldsymbol{q}-\boldsymbol{k},\eta\right)^{\textnormal{av}} & \simeq & P_{L}\!\left(q,\eta\right)\frac{k^{2}}{\pi^{2}} \biggl[ \left(\alpha^{\textnormal{SPT}}\, k\,\partial_{k}+\beta^{\textnormal{SPT}}\right)P_{L}\!\left(k,\eta\right) \! \int\! dl\, l^{2}\left(\frac{P_{L}\!\left(l,\eta\right)}{l^{2}}\right) \nn \\
 &  & \qquad \qquad \quad \;+\gamma^{\textnormal{SPT}} \,k^{2} \!
\int\! dl\, l^{2}\left(\frac{P_{L}^{2}\!\left(l,\eta\right)}{l^{4}}\right) \biggr]\,,
\end{eqnarray}
where we defined 
\be
\beta^{\textnormal{SPT}}=\beta_{1}^{\textnormal{SPT}}+\beta_{2}^{\textnormal{SPT}}=-\frac{3719}{13230}\,,
\ee
and
\be
\alpha^{\textnormal{SPT}}=\alpha_{1}^{\textnormal{SPT}}+\alpha_{2}^{\textnormal{SPT}}=\frac{61}{1890}\,.
\ee


\newpage

\begin{thebibliography}{10}

\bibitem{Maldacena:2011nz}
  J.~M.~Maldacena and G.~L.~Pimentel,
  ``On graviton non-Gaussianities during inflation,''
  JHEP {\bf 1109} (2011) 045
  [arXiv:1104.2846 [hep-th]].


\bibitem{baldauf2}
T.~Baldauf, U.~Seljak, L.~Senatore and M.~Zaldarriaga,
  ``Galaxy Bias and non-Linear Structure Formation in General Relativity,''
  JCAP {\bf 1110}, 031 (2011)
  [arXiv:1106.5507 [astro-ph.CO]].


\bibitem{Sherwin}
  B.~D.~Sherwin and M.~Zaldarriaga,
  ``The Shift of the Baryon Acoustic Oscillation Scale: A Simple Physical Picture,''
  Phys.\ Rev.\ D {\bf 85} (2012) 103523
  [arXiv:1202.3998 [astro-ph.CO]].



\bibitem{Kehagias:2013yd} 
  A.~Kehagias and A.~Riotto,
  ``Symmetries and Consistency Relations in the Large Scale Structure of the Universe,''
  Nucl.\ Phys.\ B {\bf 873}, 514 (2013)
  [arXiv:1302.0130 [astro-ph.CO]].


\bibitem{Peloso:2013zw}
  M.~Peloso and M.~Pietroni,
  ``Galilean invariance and the consistency relation for the nonlinear squeezed bispectrum of large scale structure,''
  JCAP {\bf 1305} (2013) 031
  [arXiv:1302.0223] .


\bibitem{Creminelli:2013mca}
  P.~Creminelli, J.~Noreña, M.~Simonović and F.~Vernizzi,
  ``Single-Field Consistency Relations of Large Scale Structure,''
  JCAP {\bf 1312} (2013) 025
  [arXiv:1309.3557 [astro-ph.CO]].

\bibitem{Peloso:2013spa}
  M.~Peloso and M.~Pietroni,
  ``Ward identities and consistency relations for the large scale structure with multiple species,''
  JCAP {\bf 1404} (2014) 011
  [arXiv:1310.7915 [astro-ph.CO]].

\bibitem{Creminelli:2013poa}
  P.~Creminelli, J.~Gleyzes, M.~Simonović and F.~Vernizzi,
  ``Single-Field Consistency Relations of Large Scale Structure. Part II: Resummation and Redshift Space,''
  JCAP {\bf 1402} (2014) 051
  [arXiv:1311.0290 [astro-ph.CO]].


\bibitem{Creminelli:2013nua}
  P.~Creminelli, J.~Gleyzes, L.~Hui, M.~Simonović and F.~Vernizzi,
  ``Single-Field Consistency Relations of Large Scale Structure. Part III: Test of the Equivalence Principle,''
  JCAP {\bf 1406} (2014) 009
  [arXiv:1312.6074 [astro-ph.CO]].

\bibitem{paolo1}
P.~Creminelli and M.~Zaldarriaga,
  ``Single field consistency relation for the 3-point function,''
  JCAP {\bf 0410}, 006 (2004)
  [astro-ph/0407059].
  
  \bibitem{dansoft} V.~Assassi, D.~Baumann and D.~Green,
  ``On Soft Limits of Inflationary Correlation Functions,''
  JCAP {\bf 1211}, 047 (2012)
  [arXiv:1204.4207 [hep-th]].

\bibitem{Gold} 
  W.~D.~Goldberger, L.~Hui and A.~Nicolis,
  ``One-particle-irreducible consistency relations for cosmological perturbations,''
  Phys.\ Rev.\ D {\bf 87}, no. 10, 103520 (2013)
  [arXiv:1303.1193 [hep-th]].
\bibitem{fgp} R. Flauger, D. Green and R. A. Porto, ``On squeezed limits in single field inflation: Part I,"
JCAP {\bf 1308} (2013) 032
[arXiv:1303.1430 [hep-th]]


\bibitem{Lam}
K.~Hinterbichler, L.~Hui and J.~Khoury,
  ``An Infinite Set of Ward Identities for Adiabatic Modes in Cosmology,''
  JCAP {\bf 1401}, 039 (2014)
  [arXiv:1304.5527 [hep-th]].

\bibitem{Creminelli:2013cga}
  P.~Creminelli, A.~Perko, L.~Senatore, M.~Simonović and G.~Trevisan,
  ``The Physical Squeezed Limit: Consistency Relations at Order $q^2$,''
  JCAP {\bf 1311} (2013) 015
  [arXiv:1307.0503 [astro-ph.CO]].

\bibitem{Valageas:2013zda}
  P.~Valageas,
  ``Angular averaged consistency relations of large-scale structures,''
  arXiv:1311.4286~.


\bibitem{Kehagias:2013paa}
  A.~Kehagias, H.~Perrier and A.~Riotto,
  ``Equal-time Consistency Relations in the Large-Scale Structure of the Universe,''
  arXiv:1311.5524 [astro-ph.CO].
  
\bibitem{Scoccimarro:2000zr} 
  R.~Scoccimarro,
  ``A new angle on gravitational clustering,''
  astro-ph/0008277.

\bibitem{Bernardeau:2001qr}
  F.~Bernardeau, S.~Colombi, E.~Gaztanaga and R.~Scoccimarro,
  ``Large scale structure of the universe and cosmological perturbation theory,''
  Phys.\ Rept.\  {\bf 367} (2002) 1
  [astro-ph/0112551].


  
\bibitem{Pietroni:2008jx}
  M.~Pietroni,
  ``Flowing with Time: a New Approach to Nonlinear Cosmological Perturbations,''
  JCAP {\bf 0810} (2008) 036
  [arXiv:0806.0971 [astro-ph]].

\bibitem{Li:2014sga} 
  Y.~Li, W.~Hu and M.~Takada,
  ``Super-Sample Covariance in Simulations,''
  Phys.\ Rev.\ D {\bf 89}, 083519 (2014)
  [arXiv:1401.0385 [astro-ph.CO]].
  
\bibitem{Wagner:2014aka} 
  C.~Wagner, F.~Schmidt, C.~T.~Chiang and E.~Komatsu,
  ``Separate Universe Simulations,''
  arXiv:1409.6294 [astro-ph.CO].
  
\bibitem{Chiang:2014oga} 
  C.~T.~Chiang, C.~Wagner, F.~Schmidt and E.~Komatsu,
  ``Position-dependent power spectrum of the large-scale structure: a novel method to measure the squeezed-limit bispectrum,''
  JCAP {\bf 1405}, 048 (2014)
  [arXiv:1403.3411 [astro-ph.CO]].


\bibitem{eft1} 
  D.~Baumann, A.~Nicolis, L.~Senatore and M.~Zaldarriaga,
  ``Cosmological Non-Linearities as an Effective Fluid,''
  JCAP {\bf 1207}, 051 (2012)
  [arXiv:1004.2488 [astro-ph.CO]].

\bibitem{eft2} J.~J.~M.~Carrasco, M.~P.~Hertzberg and L.~Senatore,
  ``The Effective Field Theory of Cosmological Large Scale Structures,''
  JHEP {\bf 1209}, 082 (2012)
  [arXiv:1206.2926 [astro-ph.CO]].
  
\bibitem{Jain:1995kx} 
  B.~Jain and E.~Bertschinger,
  ``Selfsimilar evolution of cosmological density fluctuations,''
  Astrophys.\ J.\  {\bf 456} (1996) 43
  [astro-ph/9503025].

\bibitem{Bernardeau:2011vy}
  F.~Bernardeau, N.~Van de Rijt and F.~Vernizzi,
  ``Resummed propagators in multi-component cosmic fluids with the eikonal approximation,''
  Phys.\ Rev.\ D {\bf 85} (2012) 063509
  [arXiv:1109.3400].

\bibitem{Blas:2013bpa} 
  D.~Blas, M.~Garny and T.~Konstandin,
  ``On the non-linear scale of   cosmological perturbation theory,''
  JCAP {\bf 1309} (2013) 024
  [arXiv:1304.1546 [astro-ph.CO]].

\bibitem{Sugiyama:2013gza}
  N.~S.~Sugiyama and D.~N.~Spergel,
  ``How does non-linear dynamics affect the baryon acoustic oscillation?,''
  JCAP {\bf 1402} (2014) 042
  [arXiv:1306.6660 [astro-ph.CO]].

\bibitem{left} R.~A.~Porto, L.~Senatore and M.~Zaldarriaga,
  ``The Lagrangian-space Effective Field Theory of Large Scale Structures,''
  JCAP {\bf 1405}, 022 (2014)
  [arXiv:1311.2168 [astro-ph.CO]].

\bibitem{Carrasco:2013sva}
  J.~J.~M.~Carrasco, S.~Foreman, D.~Green and L.~Senatore,
  ``The 2-loop matter power spectrum and the IR-safe integrand,''
  arXiv:1304.4946 [astro-ph.CO].

 \bibitem{Dan} J.~J.~M.~Carrasco, S.~Foreman, D.~Green and L.~Senatore,
  ``The Effective Field Theory of Large Scale Structures at Two Loops,''
  JCAP {\bf 1407}, 057 (2014)
  [arXiv:1310.0464 [astro-ph.CO]].

\bibitem{Pajer}  
  L.~Mercolli and E.~Pajer,
  ``On the velocity in the  Effective Field Theory of Large Scale Structures,''
  JCAP {\bf 1403}, 006 (2014)
  [arXiv:1307.3220 [astro-ph.CO]].
  
\bibitem{Baldauf:2014qfa}
  T.~Baldauf, L.~Mercolli, M.~Mirbabayi and E.~Pajer,
  ``The Bispectrum in the Effective Field Theory of Large Scale Structure,''
  arXiv:1406.4135 [astro-ph.CO].

\bibitem{Leo}
 R.~E.~Angulo, S.~Foreman, M.~Schmittfull and L.~Senatore,
  ``The One-Loop Matter Bispectrum in the Effective Field Theory of Large Scale Structures,''
  arXiv:1406.4143 [astro-ph.CO].
  
\bibitem{Audren:2011ne}
  B.~Audren and J.~Lesgourgues,
  ``Non-linear matter power spectrum from Time Renormalisation Group: efficient computation and comparison with one-loop,''
  JCAP {\bf 1110} (2011) 037
  [arXiv:1106.2607 [astro-ph.CO]].

\bibitem{Juergens:2012ap} 
  G.~Juergens and M.~Bartelmann,
  ``Perturbation Theory Trispectrum in the Time Renormalisation Approach,''
  arXiv:1204.6524 [astro-ph.CO].

\bibitem{Zeldovich:1969sb} 
  Y.~B.~Zeldovich,
  ``Gravitational instability: An Approximate theory for large density perturbations,''
  Astron.\ Astrophys.\  {\bf 5}, 84 (1970).
  
  
\bibitem{checkkprv} T.~Nishimichi and P.~Valageas,
  ``Testing the equal-time angular-averaged consistency relation of the gravitational dynamics in N-body simulations,''
  Phys.\ Rev.\ D {\bf 90}, 023546 (2014)
  [arXiv:1402.3293].
  
\end{thebibliography}
\end{document}